\documentclass[pra,twocolumn,preprintnumbers,amsmath,amssymb,superscriptaddress,showpacs]{revtex4-1}
\usepackage{graphicx}
\usepackage{latexsym}
\usepackage{amsmath}
\usepackage{graphics}
\usepackage{amssymb}
\usepackage{layout}
\usepackage{verbatim}
\usepackage{amsfonts,epsfig}
\usepackage{natbib}
\usepackage{hyperref}

\newcommand{\beq}{\begin{equation}}
\newcommand{\eeq}{\end{equation}}
\newcommand{\bea}{\begin{eqnarray}}
\newcommand{\eea}{\end{eqnarray}}

\newcommand{\mean}[1]{\langle{#1}\rangle{}}

\newcommand{\dxi}{\delta\xi}
\newcommand{\te}{t}  
\newcommand{\vq}{v_{2}}

\begin{document}

\title{Dynamical decoupling noise spectroscopy at an optimal working point of a qubit}
\author{{\L}ukasz Cywi{\'n}ski}
\email{lcyw@ifpan.edu.pl}
\affiliation{Institute of Physics, Polish Academy of Sciences, al.~Lotnik{\'o}w 32/46, PL 02-668 Warszawa, Poland}

\date{\today}

\begin{abstract}
I present a theory of environmental noise spectroscopy via dynamical decoupling of a qubit at an optimal working point. Considering a sequence of $n$ pulses and pure dephasing due to quadratic coupling to Gaussian distributed noise $\xi(t)$, 
I use the linked-cluster (cumulant) expansion to calculate the coherence decay. Solutions allowing for reconstruction of spectral density of noise are given. For noise with correlation time shorter than the timescale on which coherence decays, the noise filtered by the dynamical decoupling procedure can be treated as effectively Gaussian at large $n$, and well-established methods of noise spectroscopy can be used to reconstruct the spectrum of $\xi^{2}(t)$ noise. On the other hand, for noise of dominant low-frequency character ($1/f^{\beta}$ noise with $\beta \! > \! 1$), an infinite-order resummation of the cumulant expansion is necessary, and it leads to an analytical formula for coherence decay having a power-law tail at long times. In this case, the coherence at time $\te$ depends both on spectral density of $\xi(t)$ noise at $\omega \! = \! n\pi/t$, and on the effective low-frequency cutoff of the noise spectrum, which is typically given by the inverse of the data acquisition time. 
Simulations of decoherence due to purely transverse noise show that the analytical formulas derived in this paper apply in this often encountered case of an optimal working point, provided that the number of pulses is not very large, and that the longitudinal qubit splitting is much larger than the transverse noise amplitude.
\end{abstract}


\maketitle

\section{Introduction}
Interaction with the environment leads to decoherence \cite{Zurek_RMP03} of quantum states of small systems,  such as qubits. While it is an effect which should be avoided in the context of quantum computation (\emph{e.g.} by performing gate operations on timescales much shorter than the decoherence time, and by using error correction \cite{Nielsen_Chuang}), in other contexts the phenomenon is interesting in itself \cite{Zurek_RMP03}. Recently there has been a lot of attention devoted to using measurements of a qubit's coherence dynamics to acquire information on the environmental noise affecting the qubit. While the measurements of a qubit's energy relaxation give information on high frequency ($\omega \! > \! k_{\text{B}}T$) quantum noise \cite{Schoelkopf_spectrometer,Astafiev_PRL04}, the qubit's dephasing is sensitive to low-frequency environmental fluctuations, which often can be assumed to be classical. The measurement of dephasing during the free evolution of the qubit typically provides only information about the total noise power at low frequencies, since most often the average over many repetitions of an experiment leads to an apparent decay of the signal due to inhomogeneous broadening, i.e.~the observed dephasing is dominated by the slowest environmental fluctuations \cite{Makhlin_CP04,Ithier_PRB05}.
While the possibility of characterizing the environmental noise by analysis of the qubit's coherence decay was the subject of many investigations focusing on specific kinds of environments \cite{Falci_PRA04,Faoro_PRL04,Benedetti_PRA14,Kotler_Nature11}, 
recently noise spectroscopy methods based on dynamical decoupling (DD) \cite{Viola_PRA98,Uhrig_PRL07,Khodjasteh_PRA07,Cywinski_PRB08,Biercuk_JPB11,Yang_FP11,Khodjasteh_NC13} of the qubit from its environment have become widely used in experiments \cite{Biercuk_Nature09,Bylander_NP11,Alvarez_PRL11,Yuge_PRL11,Medford_PRL12,Dial_PRL13,Staudacher_Science13,Muhonen_arXiv14,Romach_arXiv14}. 
The application of multiple short pulses rotating the qubit's state removes the influence of the quasi-static fluctuations, and in fact for a large number of appropriately spaced pulses it can filter out the noise at all frequencies with exception of a set of narrow-band ranges \cite{Kotler_Nature11,Alvarez_PRL11,Bylander_NP11}, the contributions from which determine the time-dependence of coherence decay. It is thus possible to turn the qubit into a true \textit{spectrometer of noise} \cite{Biercuk_Nature09,Bylander_NP11,Alvarez_PRL11,Yuge_PRL11,Medford_PRL12,Dial_PRL13,Staudacher_Science13,Muhonen_arXiv14,Romach_arXiv14}.

Various schemes of noise spectroscopy with qubits (including the ones not using DD \cite{Yan_PRB12,Fink_PRL13})  have been mainly applied to the case of pure dephasing due to linear coupling to the noise $\xi(t)$, \textit{i.e.}~for the qubit-environment interaction of the $v_{1}\hat{\sigma}_{z}\xi(t)$ form. 
In such a case only the off-diagonal element of the qubit's density matrix, $\rho_{+-}(t)$, decays. 
Under the common assumption of Gaussian statistics of $\xi(t)$, the noise is fully characterized by its spectral density $S(\omega)$, which is the Fourier transform of its two-point correlation function $C(t) \! = \! \mean{\xi(t)\xi(0)} - \mean{\xi}^2$, where $\mean{...}$ denotes the averaging over the realizations of the stochastic process.
The noise spectroscopy methods which are the most relevant here are based on the fact that  $\rho_{+-}(t)$ is given in this case by an expression containing an integral of $S(\omega)$ multiplied by a sequence-specific \emph{filter function} \cite{deSousa_TAP09,Cywinski_PRB08,Biercuk_Nature09,Biercuk_JPB11}. 
For large number $n$ of pulses in an appropriately chosen DD sequence (i.e.~a Carr-Purcell \cite{Carr_Purcell} sequence), we have $\rho_{+-}(t) \! \propto \! \exp[-\chi^{l}_{2}(t)]$ with $\chi^{l}_{2}(t) \! \sim \! t S(n\pi/t)$. Thus, by changing $n$ and $\te$, $S(\omega)$ can be reconstructed \cite{Alvarez_PRL11,Yuge_PRL11,Staudacher_Science13,Muhonen_arXiv14,Romach_arXiv14} from measurements of $\rho_{+-}(t)$.  

However, one often encounters the case in which the coupling to the noise is quadratic:
\beq
\hat{H} = \frac{1}{2}[\Omega + \vq\xi^{2}(t')]\hat{\sigma}_{z} \,\, , \label{eq:H}
\eeq
where $\Omega$ is the controlled qubit splitting, and $\vq$ is the coupling constant. Obtaining such a form of qubit-noise coupling usually requires tuning some parameters of the qubit to specific values, and the point in the parameter space in which Eq.~(\ref{eq:H}) holds exactly (or, as will be discussed later, approximately), is usually referred to as the Optimal Working Point (OWP) of the qubit. 
At such an OWP the influence of noise is typically suppressed compared to the linear coupling case, and the qubit dephasing time is thus enhanced. If we were able to perform noise spectroscopy at the OWP, we would  gain information about noise in a wider range of frequencies, since with longer dephasing time it is possible to acquire good-quality data on a longer timescale, which should give access to more detailed information about noise spectrum at lower frequencies. 
We cannot however use the simple formulas connecting $\rho_{+-}(t)$ to $S(\omega)$ given before: even though $\xi(t)$ is assumed to have Gaussian statistics, its square $\xi^{2}(t)$ is \textit{not} Gaussian-distributed. The derivation of useful formulas for coherence decay under DD at the OWP is the goal of this paper.

Since their application for magnetic-field tuning of transitions used in atomic clocks \cite{Bollinger_PRL85}, the OWPs have been identified for almost all kinds of qubits, including superconducting charge \cite{Vion_Science02,Ithier_PRB05} and flux \cite{Yoshihara_PRL06,Kakuyanagi_PRL07} qubits, semiconductor quantum dot (QD) charge qubits \cite{Petersson_PRL10}, mixed electronic-nuclear spin qubits based on electrons bound to bismuth donors in silicon \cite{Wolfowicz_NN13,Balian_PRB14}, 
and a recently realized ``resonant exchange qubit'' based on a triple QD \cite{Medford_NN13,Medford_PRL13}. The existence of the OWP with respect to charge noise was also predicted theoretically for singlet-triplet qubits in double QDs \cite{Stopa_NL08,Li_PRB10,Ramon_PRB10}. The most relevant noises are the $1/f^{\beta}$ charge and flux noise ubiquitous in condensed matter \cite{Paladino_RMP14} and random telegraph noise (RTN) coming from a two-level fluctuator strongly coupled to the qubit \cite{Galperin_PRL06,Bergli_PRB06,Bergli_PRB07,Ramon_PRB12}. 
Furthermore, in the recently investigated case of an OWP of a spin qubit in Bi-doped silicon \cite{Wolfowicz_NN13,Balian_PRB14}, the Ornstein-Uhlenbeck (OU) noise will become relevant for isotopically purified samples, in which the electron coherence will be limited by interaction with other electron spins, not the nuclear spins. In this case the influence of the dipolarly coupled electron spin bath can be mapped on interaction with OU noise \cite{Dobrovitski_PRL09,deLange_Science10}. 

The performance of DD protocols at an OWP (or close to it) was most extensively discussed for superconducting qubits (and double quantum dot qubits sensitive to charge noise), for which either RTN \cite{Bergli_PRB07,Ramon_PRB12}, or a $1/f$ type noise (due to many RTN sources, often - but not always - well approximated by a Gaussian process) was considered \cite{Falci_PRA04,Faoro_PRL04}. However, since these papers focused either on non-Gaussian noise, or on the presence of signatures of non-Gaussianity of $1/f$ noise and issues specific to the bath consisting of two-level fluctuators, there is a need for development of an easy-to-use method of noise spectroscopy for the case of Gaussian noise at an OWP.

Here I investigate the possibility of using DD sequences for performing spectroscopy of Gaussian noise $\xi(t')$ at the OWP. In order to deal with non-Gaussian statistics of $\xi^{2}(t')$ I use linked-cluster (cumulant) expansion to average the qubit's phase over realizations of noise, building on the seminal papers \cite{Makhlin_PRL04,Makhlin_CP04,Falci_PRL05} in which free evolution dephasing at an OWP was considered. Potentially useful solutions allowing for noise spectroscopy appear in two cases. 

The first case is that of noise with non-singular spectrum at low frequencies, i.e.~of noise with finite autocorrelation time. I will argue that at large $n$ the dephasing at relatively short timescales can then be described in a way similar to the linear-coupling case, only with the spectral density of the $\xi^{2}(t')$ process, $S_{2}(\omega)$, appearing in the well-known formulas \cite{deSousa_TAP09,Cywinski_PRB08}. Thus, the experimentally established methods of noise spectroscopy \cite{Bylander_NP11,Medford_PRL12} can be used to reconstruct $S_{2}(\omega)$. This result can be explained by the effective ``Gaussianization'' of the noise experienced by the qubit subjected to many $\pi$ pulses. Formally, the second term in the cumulant expansion of $\rho_{+-}(t)$ becomes a good approximation to an exact result (up to a time comparable to $T_{2}$) for large enough $n$.

The second case is that of $1/f^{\beta}$ noise with $\beta \! > \! 1$ at low frequencies. The noise autocorrelation time $t_{c}$ is then ill-defined, or its value is simply irrelevant when the qubit evolution time $\te$ is much shorter than $t_{c}$. Furthermore, for the $\beta \! > \! 1$ case which we consider, the noise power is concentrated at low frequencies. The solution is obtained by separate averaging over fast ($\omega \! > \! 1/t$) and slow ($\omega_{0} \! < \! \omega \! < \! 1/\te$, where $\omega_{0}$ is the low-frequency cutoff) fluctuations. Such an approximate procedure allows for a resummation of cumulant expansion and derivation of closed formulas for $\rho_{+-}(t)$.  
The coherence at time $t$ under a sequence of many pulses is shown to be determined by $S(n\pi/t)$ (similarly to the linear coupling case), thus allowing for reconstruction of $S(\omega)$.
The characteristic feature of this solution is an appearance of a power-law tail of the coherence signal.  
Interestingly, for noise with no intrinsic low-frequency cutoff $\omega_{0}$, the coherence time $T_{2}$ scales with both $n$, and the total measurement time $T_{M}$ which determines the effective cutoff $\omega_{0} \! \sim \! 1/T_{M}$.

It is important to note here that the OWP Hamiltonian given in Eq.~(\ref{eq:H}) 
often appears as an effective approximate Hamiltonian when the qubit having large longitudinal splitting, $\Omega$, is exposed to transverse noise $v_{t}\hat{\sigma}_{x}\xi(t)$. While the domain of approximate equivalence of the calculation using the transverse noise and the one employing the effective Hamiltonian can be easily established in the case of free evolution, it is less clear on what timescale one can use the latter approach when the qubit is subjected to a DD sequence. This is mostly related to accumulation of errors caused by interplay of many pulses (along specific axes) and the random tilts of the qubit quantization axis caused by the transverse noise \cite{Mkhitaryan_PRB14}.
This effect is not accounted for in the effective Hamiltonian treatment. 
The goal of this paper is to address a theoretical question of derivation of analytical (and potentially useful for noise spectroscopy purposes) formulas for time-dependence of coherence when the qubit-noise coupling is given by Eq.~(\ref{eq:H}). 
While the example results given in the paper are shown to agree with simulations of decoherence due to transverse noise in the presence of large enough $\Omega$ splitting, the question of the precise extent of the domain (defined by $n$ and the timescale of interest) of quantitative applicability of the presented theory to specific cases of OWPs occurring due to presence of transverse noise remains to be further investigated. 
However, let me note that in experimental implementations of DD based noise spectroscopy the issue of accumulation of errors of realistic pulses is always present, 
and in practical cases the reliable results are always confined to the regime of not-very-large numbers of pulses. When working at the OWP, it is crucial to have a qualitative picture of expected coherence decay under DD for relatively small number of pulses, and the calculations based on Eq.~(\ref{eq:H}) given below provide precisely such kind of picture.

The paper is organized in the following way. 
In Section \ref{sec:OWP}  the basic general facts about the Optimal Working Point of the qubit and the types of noise most relevant for qubits are reviewed. 
Section \ref{sec:linear} contains an outline of derivation of noise spectroscopy formulas for the case of linear coupling to noise. This is given here in order to make the paper self-contained and to establish the reference point for the OWP noise spectroscopy methods discussed later. In Section \ref{sec:LCE} I present a general form of the solution for coherence dynamics at an OWP (assuming an effective Hamiltonian description) using the cumulant expansion, and the short-time behavior of this solution is discussed in Section \ref{sec:short}. Approximate analytical solutions for $\rho_{+-}(t)$ at longer times are then given in Section \ref{sec:Gaussian} and \ref{sec:1f} for cases of noise with finite autocorrelation time and of $1/f^{\beta}$ noise, respectively. The analytical results given in these sections are compared with numerical simulations employing Ornstein-Uhlenbeck noise with autocorrelation time $t_{c}$. By varying the coupling to the noise (which determines the timescale $T_{2}$ at which coherence is non-negligible) we can illustrate both the regime of $T_{2} \! \gg \! t_{c}$ (for weak coupling), and the case of $T_{2} \! \ll \! t_{c}$ (for strong coupling), which effectively corresponds to $1/\omega^{2}$ noise.
The relation between the results obtained with the effective Hamiltonian and a model in which the OWP arises due to transverse coupling to noise in the presence of large longitudinal splitting is discussed in Section \ref{sec:transverse}.

\section{Optimal working point and the noise affecting the qubit} \label{sec:OWP}
Quadratic qubit-noise coupling given in Eq.~(\ref{eq:H}) can appear in two ways. In the first one (the ``intrinsic'' OWP) we have the Hamiltonian
\beq
\hat{H} = \frac{1}{2}\left[ \Omega' + \delta \Omega(\xi_{0}+\xi(t)) \right ] \hat{\sigma}_{z} \,\, ,
\eeq
where $\delta \Omega$ is the energy offset which depends on a controlled parameter $\xi_{0}$ and a stochastic variable $\xi(t)$, and we assume that $\partial \delta \Omega(x)/\partial x|_{x=\xi_{0}} \! = \! 0$. In the lowest nonvanishing order with respect to $\xi(t)$ we obtain the Hamiltonian from Eq.~(\ref{eq:H}) with $\Omega\! = \! \Omega' + \delta\Omega(\xi_{0})$ and  $v_{2} \! = \! \frac{1}{2}\partial^{2}\delta \Omega/\partial x^{2}|_{x=\xi_{0}}$. 
Such an \emph{intrinsic} OWP appears for example in superconducting qubits of the types which are mostly affected by flux noise \cite{Shnirman_PS02,Yoshihara_PRL06,Kakuyanagi_PRL07} - the nonlinear dependence of the Josephson energy on the magnetic flux allows for the presence of an extremum of $\delta \Omega(\xi)$ dependence.

The second type of the OWP (the \emph{extrinsic} one) appears when the Hamiltonian is given by
\beq
\hat{H} = \frac{\Omega}{2}\hat{\sigma}_{z} + v_{t}\frac{\xi(t')}{2}\hat{\sigma}_{x} \,\, .  \label{eq:Ht}
\eeq
This is the case in which the qubit, having energy eigenstates quantized along the $z$ axis, is exposed to transverse noise. The noise is assumed above to be along the $x$ direction only, since this is the often encountered case \cite{Vion_Science02,Ithier_PRB05,Shnirman_PS02,Bergli_PRB06,Bergli_PRB07,Medford_PRL13}. 
When the characteristic energy scale of the noise, $v_{t}\sigma$, with $\sigma^{2} \! \equiv \! \mean{\xi^{2}}$, is much smaller than $\Omega$, in the lowest orders of expansion in $v_{t}\sigma/\Omega$ the noise is causing the tilting of the qubit's quantization axis by angle $\approx v_{t}\sigma/\Omega$ and a fluctuation of the qubit's precession frequency around this axis of magnitude $\approx v^{2}_{t}\sigma^2/2\Omega$. This geometric interpretation of the qubit's evolution should make it clear that in the process of dephasing of a freely evolving qubit (i.e.~the decoherence of a superposition of eigenstates of $\hat{\sigma}_{z}$ without any additional pulses rotating the qubit's state) the effect of noise on precession frequency is dominant, while the axis tilting contributes only a small (on the order of $v^{2}_{t}\sigma^{2}/\Omega^2$) correction to the coherence signal magnitude. The effective pure dephasing Hamiltonian is again of the form given in Eq.~(\ref{eq:H}), only with $v_{2} \! = \! v^{2}_{t}/2\Omega$.

It is  important to note that this approach should be used with caution in the case of decoherence of the qubit exposed to multiple pulses (rotations about $x$ or $y$ axes). Firstly, in the presence of the uniaxial transverse noise a difference in performance of DD pulses ($\pi$ rotations) along the $x$ and $y$ axes is expected \cite{Bergli_PRB07,Ramon_PRB12}. Secondly, these $\pi$ rotations become imperfect in the presence of transverse noise, and the resulting pulse errors can accumulate with increasing $n$ \cite{Mkhitaryan_PRB14}. Nevertheless, in the following Sections I will develop a theory of DD noise spectroscopy at an OWP using the Hamiltonian from Eq.~(\ref{eq:H}), and the issue of accuracy of these calculations when the original Hamiltonian is in fact given by Eq.~(\ref{eq:Ht}) will be revisited in Section \ref{sec:transverse}.  

The stochastic process $\xi(t')$ will be assumed to be stationary and Gaussian. It should be noted that the random telegraph noise (RTN), considered in other works in the context of extrinsic OWP \cite{Bergli_PRB06,Bergli_PRB07,Ramon_PRB12}, is thus excluded from the following considerations, since it has non-Gaussian statistics.

\section{Noise spectroscopy for linear coupling to Gaussian dephasing noise} \label{sec:linear}
Let us recount here the results for dephasing due to $v_{1}\xi(t')\hat{\sigma}_{z}$ coupling to Gaussian noise \cite{deSousa_TAP09,Cywinski_PRB08}. We focus now on evolution of a superposition of the two states of the qubit under the application of $n$ ideal $\pi$ pulses ($\pi$ rotations about $x$ or $y$ axis) applied in the Carr-Purcell (CP) sequence \cite{Carr_Purcell}, \textit{i.e} with pulses at times $\tau_{k} \! = \! (k-\frac{1}{2})\te/n$, with $\te$ being the total sequence time. The CP sequence is chosen because its use leads to a particularly transparent method of noise spectroscopy at large $n$. It is also simple to implement, and in the realistic case of imperfect pulses, a simple choice of the initial pulse axis vs the axis used for the subsequent rotations introduced by Meiboom and Gill results in the CPMG protocol which is quite robust against systematic pulse errors \cite{Meiboom_Gill}.
We define the decoherence function
\beq
W_{l}(t) =  \left\langle \exp\left( -iv_{1}\int_{0}^{\te}f_{t}(t') \xi(t')\text{d}t' \right ) \right\rangle \,\, , \label{eq:Wl}
\eeq
where $f_{t}(t')$ is a temporal filter function, which is zero for $t'$ outside of the $[0,\te]$ range, and equal to $\pm \! 1$ within it, changing sign at each pulse time $\tau_{k}$. For an even number of pulses $n$ the decoherence function is related to the off-diagonal element of the density matrix by $W(\te) \!= \! \rho_{+-}(\te)/\rho_{+-}(0)$, while for odd $n$ we should replace $\rho_{+-}(0)$ by $\rho_{-+}(0)$ in this formula. 

A simple Gaussian average gives a closed result for the decoherence function in this case: $W_{l}(\te) \! = \! \exp[-\chi^{l}_{2}(\te)]$, with $\chi^{l}_{2}(\te)\! \equiv \! \frac{v^{2}_{1}}{2}R^{l}_{2}(\te)$, and 
\beq
R^{l}_{2} = 
\int_{0}^{\infty} |\tilde{f}_{t}(\omega)|^2 S(\omega) \frac{\text{d}\omega}{\pi} \,\, , \label{eq:R2ldef}
\eeq
where $\tilde{f}_{t}(\omega)$ is the Fourier transform of $f_{t}(t')$. It is useful to note now that this result can be viewed as a particularly simple case of application of the cumulant expansion technique \cite{Kubo_JPSJ62} to performing the noise average. Due to the Gaussian statistics of $\xi$, the average of $\mean{\exp(-i\Phi(t))}$ (with the phase $\Phi(t)$ defined in Eq.~(\ref{eq:Wl})) is simply given by the exponent of the second cumulant: $W_{l}(\te)\! = \! \exp(-\frac{1}{2}\mean{\Phi(t)^{2}})$. 

\begin{figure}
\includegraphics[width=\linewidth]{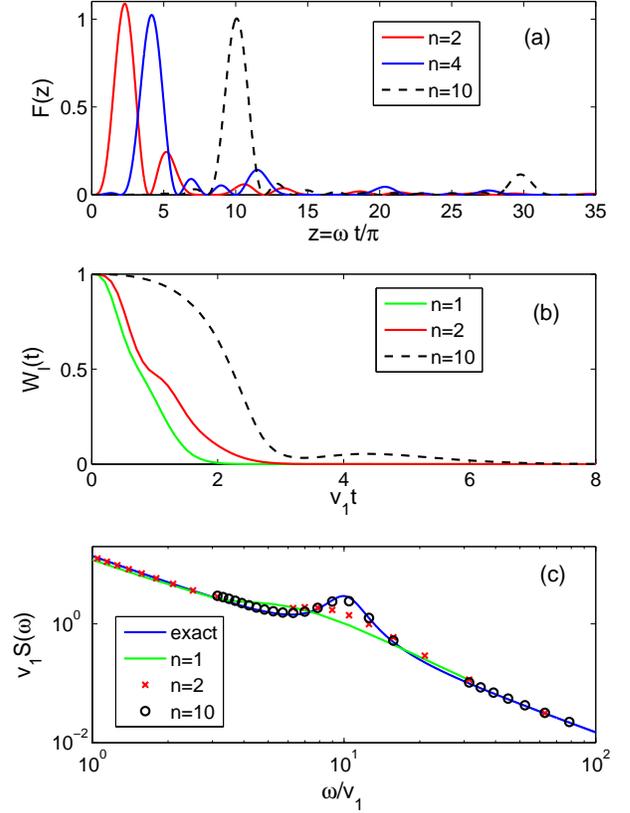}
\caption{(Color online)~(a) Frequency-domain filter function for CP sequence with $n\! = \! 2$, $4$, and $10$ pulses. The plotted function $F(z) \! = \! \pi^{2}|\tilde{f}_{t}(z)|^{2}/4\te^{2}$ with $z \! =\! \omega\te/\pi$, has a dominant peak at $z_{1} \! \approx \! n$, and the next peak at $z_{2}\! \approx \! 3n$ is $\approx \! 9$ times smaller.
(b) Calculation of $W_{l}(\te)$ for linear coupling $v_{1}\xi(t')\hat{\sigma}_{z}$ to noise with spectral density $v_{1}S(\tilde{\omega})\! = \! A/\tilde{\omega}^{3/2} + B/[\gamma^2 + (\tilde{\omega}-\omega_{p})^2]$ where $\tilde{\omega}\! = \! \omega/v_{1}$, and $A \! = \! \pi^{7/2}/4$, $B\! = \! 10$, $\omega_{p}\! = \! 10$ are dimensionless constants. (c) Demonstration of spectroscopy in this case: the results of exact calculation of $W_{l}(\te)$ from (b) are used to reconstruct $v_{1}S(\omega)$ using Eq.~(\ref{eq:Rl}). 
} \label{fig:f}
\end{figure}

The spectroscopy of $S(\omega)$ can be achieved at large $n$ when we realize that $\tilde{f}_{t}(\omega)$ can then be approximated by a series of narrow peaks \cite{Cywinski_PRB08,Bylander_NP11,Alvarez_PRL11} - this simply comes from the fact that the influence of the noise component with frequency matching the frequency of pulse application (or its harmonics) is not suppressed by the DD sequence.
For even (e) and odd (o) $n$ we have
\begin{align}
\tilde{f}^{n=e}_{t}(\omega) & \approx \frac{e^{i\omega t/2}}{\omega} (-1)^{\frac{n}{2}+1} \sum_{k=-\infty}^{\infty}(-1)^{k}  \Delta[\omega \te - 2\pi n (k-\frac{1}{2})] \,\, , \label{eq:fe} \\
\tilde{f}^{n=o}_{t}(\omega) & \approx i\frac{e^{i\omega t/2}}{\omega} (-1)^{\frac{n-1}{2}} \sum_{k=-\infty}^{\infty}  \Delta[\omega \te - 2\pi n(k-\frac{1}{2})] \,\, , \label{eq:fo}
\end{align}
where $\Delta(x)$ can be approximated by a square peak of height $2n$ and width $2\pi /\te$ centered at $x$.
This is illustrated in Fig.~\ref{fig:f}(a). 
For $S(\omega)$ which does not have very pronounced maxima (i.e.~it is mostly a non-increasing function of $\omega$) we can safely take only the first peak \cite{Bylander_NP11}, i.e.~assume $|\tilde{f}_{t}(\omega)|^{2} \! \approx \! \delta(\omega\te - \pi n)8\pi n^{2}/\omega^2$.
Note that the amplitude of peaks in $|\tilde{f}_{t}(\omega)|^{2}$ at larger $\omega$ is smaller than that of the first one by factor of $(k+1)^2$ for the $k$-th peak (see Fig.~\ref{fig:f}a). 
The single-peak approximation leads to
\beq
R^{l}_{2}(\te) \approx \frac{8 t}{\pi^2}S\left ( \frac{\pi n}{\te}\right ) \,\, . \label{eq:Rl}
\eeq  
Thus, by changing $n$ and $\te$, $S(\omega)$ can be reconstructed \cite{Alvarez_PRL11,Yuge_PRL11,Staudacher_Science13,Muhonen_arXiv14,Romach_arXiv14} from measurements of $\rho_{+-}(t)$. 
The accuracy of this formula is shown in Fig.~\ref{fig:f}(c), where it was used to reconstruct $S(\omega)$ from $W_{l}(\te)$ calculated using the exact equations given before. The peak in $S(\omega)$ can be well-described in this case with large enough $n$, but it should be kept in mind that for strongly non-monotonic $S(\omega)$ one should consider using a more complicated spectroscopy scheme  in which the contribution of filter function peaks at higher frequencies is taken into account \cite{Alvarez_PRL11}.
Alternatively, when reconstruction of $S(\omega)$ from the above formula is unreliable, it is often more robust to extract the characteristic decay timescale $T_{2}$ from the datasets corresponding to various $n$. When the part of the noise spectrum mostly responsible for coherence decay is of $1/\omega^{\beta}$ form (as it is often the case), we have $W_{l}\! \sim \! \exp[-(\te/T_{2})^{\beta+1}]$ and the decay timescale $T_{2} \! \sim \! n^{\gamma}$ with $\gamma \! = \! \beta/(\beta+1)$. Fitting of $T_{2}$ vs $n$ dependence to a power law gives the exponent $\gamma$, and it allows for rather reliable estimation of $\beta$ \cite{Cywinski_PRB08,deLange_Science10,Medford_PRL12,Muhonen_arXiv14,Siyushev_NC14}.

\section{Linked cluster expansion for quadratic coupling to Gaussian noise} \label{sec:LCE}
The decoherence function $W(\te)$ is now given by
\beq
W(\te) = \left\langle \exp\left( -i \vq \int_{0}^{\te}f_{t}(t') \xi^{2}(t')\text{d}t' \right ) \right\rangle \,\, . \label{eq:W2}
\eeq
When dealing with pure dephasing due to classical noise, the most natural theoretical approach, especially suited for transparent investigation of non-Gaussian effects, is the linked cluster expansion (LCE) \cite{Makhlin_PRL04,Makhlin_CP04,Cywinski_PRB08}. The most general form of the linked cluster theorem \cite{Negele} states that $W(\te)$ is given by an exponent of the sum of all the \emph{linked} terms that appear in expansion of RHS of Eq.~(\ref{eq:W2}). By definition these linked terms are the ones that cannot be written as products of other expressions (each including a separate average over a few $\xi$ terms). From this description it should be clear that the linked clusters are basically cumulants of the random variable appearing in the exponent on the RHS of Eq.~(\ref{eq:W2}).

Let us take a look at the lowest-order terms in this expansion, making the usual assumption of $\mean{\xi(t')} \! =\! 0$. The first order term is 
\beq
W^{(1)}(t) = -iv_{2}\int f_{t}(t')\mean{\xi^{2}(t')}\text{d}t' =  -iv_{2}C(0)\int f_{t}(t')\text{d}t' \,\, , \label{eq:We1}
\eeq
which is equal to zero because $\int f_{t}(t')\text{d}t' \! = \!0$ for any reasonable DD sequence. However, for the purpose of quickly explaining the structure of terms appearing in the expansion of $W(t)$ let us forget about this fact for a moment (note also that the derivation below applies also to the free evolution case \cite{Makhlin_PRL04}, in which $\int f_{t}(t')\text{d}t' \! = \!t$). 

In the second order of expansion we encounter a $\mean{\xi^{2}(t_{1})\xi^{2}(t_{2})}$ average. We use now the assumption of Gaussian statistics of $\xi(t')$, which tells us how the multi-point correlation functions factorize into the two-point ones. With the notation of $\xi(t_{k})\! \equiv \! \xi_{k}$ we have
\beq
\mean{\xi_{1}\xi_{1}\xi_{2}\xi_{2}} = \mean{\xi_{1}^{2}}\mean{\xi_{2}^{2}} + 2\mean{\xi_{1}\xi_{2}}^{2} = C^{2}(0) + 2C^{2}(t_{12}) \,\, , \label{eq:xi1xi2}
\eeq
where $\xi(t_{k})\! \equiv \! \xi_{k}$, $t_{kl} \! \equiv \! t_{k}-t_{l}$, and the factor of $2$ in front of the second term comes from two possibilities of pairing of $\xi_{1}$ with $\xi_{2}$. 
We obtain then
\begin{align}
W^{(2)} & = [W^{(1)}]^2 -v^{2}_{2}\int\int f_{t}(t_{1}) f_{t}(t_{2}) C(t_{12})C(t_{21})    \text{d}t_{1}\text{d}t_{2} 
\end{align}
where we have used the fact that $C(t_{12}) \! =\! C(t_{21})$ to make the expression more symmetric. The first term on the LHS is the \emph{unlinked}, while the second is the \emph{linked} one: despite the fact that the average in Eq.~(\ref{eq:xi1xi2}) factorized into the product of two averages, the presence of integrals with respect to $t_1$ and $t_2$ variables that are interlocking these averages precludes factorization of this expression. 

In the third order the $\mean{\xi_{1}^{2}\xi_{2}^{2}\xi_{3}^{2}}$ average factorizes into a $\mean{\xi_{1}^{2}}\mean{\xi_{2}^{2}}\mean{\xi_{3}^{2}}$ term, six terms of   $\mean{\xi_{k}^{2}}\mean{\xi_{l}\xi_{m}}^{2}$ type, and eight terms of  $\mean{\xi_{k}\xi_{l}}\mean{\xi_{l}\xi_{m}}\mean{\xi_{m}\xi_{k}}$ type. The latter ones lead to linked terms in $W^{(3)}$. The pattern should be clear now: in the $k$-th order of expansion of $W(t)$ we obtain $(2k-1)!!$ terms coming from possible pairings of $\xi_{k}$ operators under average, and $2^{k-1}(k-1)!$ of these terms are linked. This gives us 
\begin{align}
W(\te) & = \exp \left [ \sum_{k=1}^\infty\frac{(-iv_{2})^k}{k}R_{k}(\te) \right ] \,\, \label{eq:WR} \\ 
& = \exp \left [-\sum_{k=1}\chi_{k}(\te)  \right ]  \,\, ,  \label{eq:Wchi}
\end{align}
with the linked cluster 
contributions
\begin{align}
\!\!\! R_{k} & = \! 2^{k-1}\int f_{t}(t_{1}) \text{d}t_{1} ... \int f_{t}(t_{k}) \text{d}t_{k} C(t_{12})...C(t_{k1}) \,\, , \label{eq:Rt}\\
& \!\!\!\!\!\!\! = 2^{k-1} \! \int \frac{\text{d}\omega_{1}...\text{d}\omega_{k}}{(2\pi)^k} S(\omega_{1})...S(\omega_{k}) \tilde{f}_{t}(\omega_{12})...\tilde{f}_{t}(\omega_{k1}) \,\, , \label{eq:Rw}
\end{align} 
where $\omega_{kl} \! \equiv \! \omega_{k}-\omega_{l}$. The above equations are generalizing the results of Ref.~\cite{Makhlin_PRL04} to the case of evolution of the qubit affected by a series of ideal $\pi$ pulses.

Note that when the number of pulses $n$ is odd, then the $R_{k}(t)$ terms with odd $k$, i.e.~the ones contributing a nontrivial  phase $\psi(\te)$ to $W(t) \! = \! |W(t)|e^{i\psi(\te)}$, are identically zero. This follows from $\tilde{f}_{t}(-\omega) \! = \! -\tilde{f}_{t}(\omega)$ for odd $n$. For even $n$ the odd-order linked clusters do not vanish, and the phase contribution $\psi(t)$ is nonzero. In the following discussion we will ignore this phase. The approximations which will be used below either imply that for even $n$ we have $\psi(t)\! \approx \! 0$ (and the appearance of nonzero phase is then one of possible signatures of breaking of certain assumptions allowing for derivation of a simple solution of the problem), or the discussion of $\psi(t)$ contribution becomes cumbersome, while the numerical simulations show that the corrections brought by this term are qualitatively unimportant. In experiments one can simply choose to work with odd $n$, or perform the measurements of the qubit state along both $x$ and $y$ axes, reconstructing both real and imaginary part of $W(\te)$ in this way. 

\section{Short time behavior}  \label{sec:short}
Under dynamical decoupling the lowest-order nonvanishing term in Eq.~(\ref{eq:WR}), $R_{2}(\te)$, can be rewritten in the form analogous to that of Eq.~(\ref{eq:R2ldef})
\beq
R_{2}(t) = \int_{0}^{\infty} |\tilde{f}_{t}(\omega)|^2 S_{2}(\omega) \frac{\text{d}\omega}{\pi} \approx \frac{8 t}{\pi^2}S_{2}\left ( \frac{\pi n}{\te}\right ) \,\, ,  \label{eq:R2}
\eeq
only with $S(\omega)$ appearing in Eq.~(\ref{eq:R2ldef}) replaced by 
\beq
S_{2}(\omega) = \int S(\omega_{1})S(\omega_{1}-\omega)\frac{\text{d}\omega_{1}}{\pi} \,\, , \label{eq:S2}
\eeq
which is the  spectral density of $\xi^{2}(t')$ process, i.e.~it is the Fourier transform of its  autocorrelation function given by
\beq
C_{\xi^2}(t') \equiv \mean{\xi^{2}(t')\xi^{2}{0}} - \mean{\xi^{2}}^2 = 2C^{2}(t') \,\, .
\eeq

At very short times we have $W(t) \! \approx \! 1-\frac{1}{2}R_{2}(t)$, so that the measurement of the initial decay gives information about $S_{2}(\omega)$ at high frequencies. However, if our goal is to reconstruct the spectrum of $\xi(t')$ process, the relation between $S_{2}(\omega)$ and $S(\omega)$ has to be discussed. The fact that $S_{2}(\omega)$ is a convolution of $S(\omega)$ with itself means that in general there is no simple relation between the two spectra at the same frequency. However, in a few often-encountered cases we can find such a relation. In this paper we will focus either on case of noise with $S(\omega) \! \sim \! 1/\omega^{\beta}$ (with $\beta \! > \! 1$ and low-frequency cutoff $\omega_{0}$) in a large range of frequencies, or on the case of OU noise with correlation time $t_{c}$, for which we have the correlation function $C^{\text{OU}}(\tau) \! = \! \sigma^{2}e^{-\tau/t_{c}}$ and the corresponding spectral density:
\beq
S^{\text{OU}}(\omega) = \frac{2\sigma^2 t_{c}}{1+\omega^{2}t^{2}_{c}} \,\, , \label{eq:SOU}
\eeq
where $\sigma^2$ is the total power of the noise. For the case of $1/\omega^{\beta}$ noise, the integral in Eq.~(\ref{eq:S2}) is dominated by contributions of regions with $|\omega_{1}|$ or $|\omega_{1}-\omega|$ close to $\omega_{0}$, and a simple calculation gives $S_{2}(\omega) \! \sim \! 1/\omega^{\beta}$. A similar situation is encountered for OU noise, for which it is easy to calculate $S_{2}(\omega)$ exactly:
\beq
S^{\text{OU}}_{2}(\omega) = \frac{8\sigma^4 t_{c}}{4+\omega^{2}t^{2}_{c}} \,\, , \label{eq:S2OU}
\eeq
where we see that the high-frequency $1/\omega^2$ behavior is inherited from $S(\omega)$. At very short $\te$ we can then expect 
\beq
W(\te) \approx 1 - (\te/T_{s})^{\beta+1} \,\, ,
\eeq
where $T_{s}$ is the characteristic parameter characterizing the intial ``decay shoulder'' of $W(\te)$, and $\beta$ characterizes the power-law decay of $S(\omega)$ at large frequencies $\approx n\pi/\te$. 

The utility of the above result is diminished by the fact that high quality data in the regime of $W(\te) \! \approx \! 1$ are often not available due to finite measurement precision. Furthermore, only the high-end part of the spectrum, which causes the decay of coherence at short times, is probed here. A spectroscopy method using the measured $W(\te)$ for $\te$ comparable and larger than $T_{2}$ (roughly defined as the half-decay time of $W(\te)$) is clearly desirable. In the previously discussed case of linear coupling to $\xi$, $R^{l}_{2}$ was the only nonvanishing term. However, $R_{k}$ with $k\! > \! 2$ are not zero now, and their contribution to $W(\te)$ has to be taken into account in order to make any statement on the form of coherence decay (and its relation to the noise spectrum) beyond the short-time limit.

\section{Noise with finite correlation time - Gaussian approximation} \label{sec:Gaussian}
It is quite intuitive that with increasing $n$ the noise affecting the qubit should become better described by the Gaussian approximation, provided that the noise has a finite correlation time $t_{c}$. 
In the case of free evolution, the phase $\phi(\te) \! = \! \int_{0}^{\te} \xi^{2}(t')\text{d}t'$ is not Gaussian-distributed except at very long $t \! \gg \! t_{c}$, for which we can argue that $\phi(\te)$ is a sum of a large number of independent contributions (each from a slice of time of width $\approx \! t_c$). However, in such a free evolution case, coherence could be practically zero at such long times.
On the other hand, in the case of DD the filtered phase, $\phi_{f}(\te) \! = \! \int f_{t}(t') \xi^{2}(t')\text{d}t'$, can be viewed as a sum over $n+1$ contributions, the signs of which are arranged in such a way that the correlated parts of subsequent contributions cancel each other when $t/n \! \ll \! t_{c}$. In other words, the terms effectively contributing to the filtered phase are weakly correlated (especially when $\te \! > \! t_{c}$), allowing for application of the Central Limit Theorem, leading to Gaussian distribution of $\phi_{f}$ at large $n$. This is	 equivalent to saying that keeping only the $R_{2}(\te)$ term becomes a good approximation. 
The above argument applies to any non-Gaussian noise with finite $t_{c}$: the fact that DD suppresses non-Gaussian features was noticed for the case of a qubit linearly coupled to the source of RTN in \cite{Cywinski_PRB08,Ramon_PRB12}. 

One thing that is crucial to note here is that the above heuristic argument fails for coupling to $\xi^{2}(t')$ noise unless $\te \! \gg \! t_{c}$. The reasons for this will become clear in the next Section.

\begin{figure}[t]
\includegraphics[width=\linewidth]{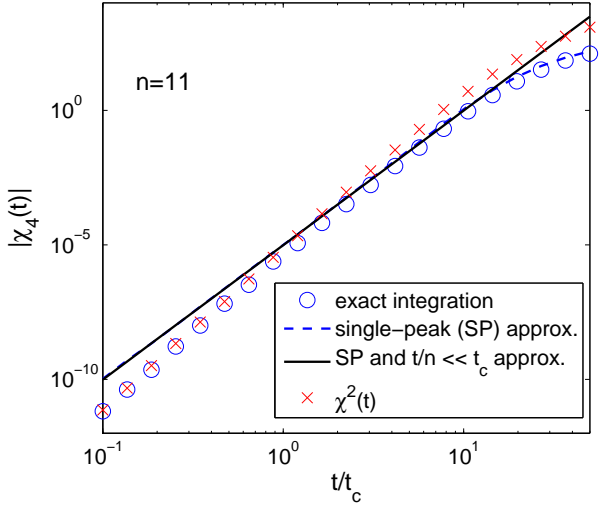}
\caption{(Color online)~Numerical calculation of $|\chi_{4}(t)|$ (circles) for Ornstein-Uhlenbeck noise with $\sigma\! = \! 1$ and $v_{2}t_{c} \! = \! 1$, done for $n\! = \! 11$ pulses of the CP sequence. The dashed line is the ``single-peak'' (SP) approximation for $|\chi_{4}| \! =\! \frac{1}{2}R_{4}$ from Eq.~(\ref{eq:R4OU}), while the solid line is the SP approximation for $t/n \! \ll \! t_{c}$ from Eq.~(\ref{eq:chi4}). The latter formula agrees quite well with the numerical calculation for $t/n \! < t_{c} \! < t$ (i.e.~for $1\! < \! t/t_{c} \! < n$). $\chi^{2}_{2}(t)$ is shown as crosses, and one can see that for $t/t_{c}\! \ll \! 1$ we have $|\chi_{4}(t)| \! \approx \! \chi^{2}_{2}(t)$, in agreement with Eq.~(\ref{eq:chi2n}) which holds in this regime.
}
 \label{fig:R2R4}
\end{figure}

Let us explore now the possibility of reaching the Gaussian limit using the OU process as an example of noise with finite $t_{c}$. $R_{2}(t)$ is then given for large $n$ by
\beq
R^{\text{OU}}_{2}(\te) \approx \frac{64 \sigma^{4}t_{c}\te^{3}}{\pi^{2}(4\te^{2} + n^{2}\pi^{2}t^{2}_{c})} \,\, ,
\eeq
where Eqs.~(\ref{eq:R2}) and (\ref{eq:S2OU}) were used. When $\te/n \! \ll \! t_{c}$ we have then in Eq.~(\ref{eq:Wchi}) the second order term 
\beq
\chi_{2}(t)  =  \frac{v^{2}_{2}}{2}R_{2}(t) =  \frac{32}{\pi^4} \frac{(\sigma^{2}v_{2}\te)^2 (\te/t_{c})}{n^2} \,\, , \label{eq:chi2}
\eeq 
while for $\te/n \! \gg \! t_{c}$ we have $\chi_{2} \! \propto \! (v_{2} \te) (v_{2}t_{c})$, which is independent of $n$. We focus now on the former regime, in which the dynamical decoupling actually leads to enhancement of coherence time. 
For noise bounded at low frequencies the main contributions to the integrals in Eq.~(\ref{eq:Rw}) come from peaks of $\tilde{f}(\omega_{kl})$. Within the single-peak approximation used before, this means that all  $\omega_{kl} \! \approx \! \pm n\pi/\te$.
For the $4$th order term using this approximation we get
\beq
R^{\text{OU}}_{4}(\te)  \approx \frac{1024t}{3\pi^4}\int \left[ S^{\text{OU}} \Big(\omega_{1} \Big)S^{\text{OU}} \Big(\omega_{1} - \frac{n\pi}{\te} \Big) \right ]^{2}  \frac{\text{d}\omega_{1}}{\pi} \,\, , \label{eq:R4OU}
\eeq
from which we get that for $\te/n \! \ll \! t_{c}$ 
\beq
|\chi_{4}(\te)| = \frac{v^{4}_{2}}{4}R_{4}(\te) \approx \frac{4096}{3\pi^8}\frac{(\sigma^{2}v_{2}\te)^{4} (\te/t_{c})}{n^{4}} \,\, . \label{eq:chi4}
\eeq
The accuracy of approximation leading to Eq.~(\ref{eq:R4OU}) is shown in Fig.~\ref{fig:R2R4}, where a numerical calculation of $\chi_{4}(\te)$ given by exact formula (\ref{eq:Rw}) is compared with the approximate calculation. 

\begin{figure}[h]
\includegraphics[width=\linewidth]{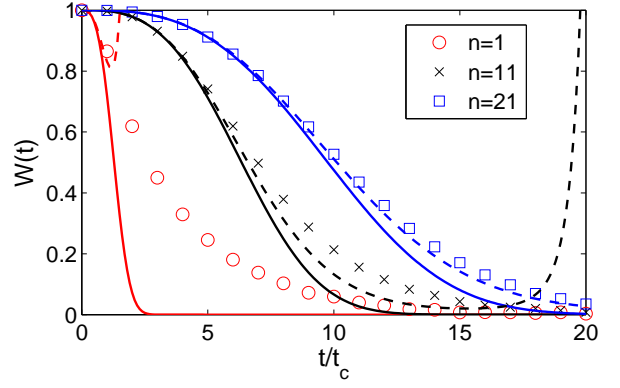}
\caption{(Color online) Decoherence function at an OWP for CP sequences with $n\!=\!1$, $11$ and $21$. Symbols are the results of numerical simulation with OU noise (with $\sigma\! = \! 1$) obtained using standard algorithms \cite{Gillespie_PRE96}. The coupling is $\vq \! = \! 1/t_{c}$. The solid lines are the Gaussian approximation $W(\te)\! =\! \exp[-\chi_{2}(t)]$, while the dashed lines are $W(\te)\! =\! \exp[-\chi_{2}(t) - \chi_{4}(t)]$ with $\chi_{2}(t)$ and $\chi_{4}(t)$ given in Eqs~(\ref{eq:chi2}) and (\ref{eq:chi4}), respectively. The divergence of the latter signifies the failure of the approach in which only a few cumulants are taken into account. Already for $n\! = \! 11$ the characteristic decay timescale $T_{2}$ is well described by the Gaussian approximation, while for $n\! = \! 21$ the two lowest cumulants are enough to describe the coherence decay by an order of magnitude from the initial value.
} \label{fig:gaussian}
\end{figure}

Let us focus now on regime of $\te \! \leq \! T_{2}$, with $T_{2}$ defined by $\chi_{2}(T_{2}) \! = \! 1$ (i.e.~$T_{2}$ is the time of decay of $W(\te)$ to $1/e$), provided that the Gaussian approximation is correct on this timescale. The condition for the latter is 
\beq
\frac{|\chi_{4}(\te)|}{\chi^{2}_{2}(\te)} \approx \frac{4t_{c}}{3\te} \ll 1 \,\, , \label{eq:g}
\eeq
where we have used Eqs.~(\ref{eq:chi2}) and (\ref{eq:chi4}). We see now that the condition for the initial decay of $W(\te)$ to be well described by Gaussian approximation  is $\te/n \! \ll \! t_c \! \ll \! \te$. At longer times, $\te \! \gg \! T_{2}$, $\chi_{2}(\te)$ becomes larger than one, and for the condition of applicability of the Gaussian approximation we can use simply $\chi_{4}(\te)\! \ll \! 1$. This leads to  $\sigma^{2}v_{2}\te/n \! \ll \! (t_{c}/\te)^{1/4}$. With $\te$ increasing to values larger that $t_{c}$, maintaining this condition at given $\te$ requires a slightly superlinear scaling of $n$ with $\te$. This is illustrated in Fig.~\ref{fig:gaussian}, where simulation of dephasing due to square of OU noise are compared with the Gaussian formula, and the next-order formula including $\chi_{4}(\te)$. For $n\! =\! 1$ the Gaussian regime does not extend to timescale when $W(\te) \! = \! 1/e$, while with $n\! =  \! 11$ and $21$ the $T_2$ time is well-described by Gaussian expressions.

While the OU noise has been used here, the physics should be qualitatively the same for any other noise with a well-defined $t_{c}$. The discussion in the next Section will make it clear that the fingerprint of the situation considered here is the disappearance of long-time power-law tail (visible for $n\!=\! 1$ in Fig.~\ref{fig:gaussian}) with increasing $n$, and enlargement of timescale on which $W(\te) \! \sim \! \exp(-(t/T_{2})^{\alpha})$. It should also be clear that the most realistic method of noise spectroscopy in this case is the investigation of $T_{2} \! \sim \! n^{\gamma}$ scaling, with $T_{2}$ times fit to the data for $W(\te)$ with the long-time tails removed.

\section{Resummation of linked clusters for low-frequency noise} \label{sec:1f} 
The calculations from the previous Section do not work for noise with an ill-defined correlation time, or when $\te \! \ll \! t_{c}$. Formally, the assumption that the peaks of $\tilde{f}(\omega_{kl})$ dominate the integrals in Eq.~(\ref{eq:Rw}) is now incorrect. The regions of very small $\omega_{k}$, in which the spectral densities diverge, also matter. Before I present an approximation for Eq.~(\ref{eq:Rw}) which works for noise with strong low-frequency component, let me give a simple qualitative explanation of the relevant physics.

The first thing to note is that for noise with a spectrum diverging at low frequencies, one has to carefully think on the meaning of $\Omega$ in Eq.~(\ref{eq:H}). This quantity consists of ``bare'' qubit splitting renormalized by the average contribution from noise, $v_{2}\mean{\xi^{2}}$. However, we should now consider more carefully the meaning of the $\mean{...}$ average. 
Practically, $\Omega$ is measured with some accuracy before the coherence measurements are begun. The cycle of the qubit's initialization, evolution, and measurement, is then repeated many times. The key observation is that for $1/f^\beta$ noise, there will be slow, essentially static on a timescale of $\te$,  fluctuations of $\xi^{2}$. During a single evolution, the noise contribution to the qubit's splitting is 
\beq
\xi^{2}(t')\! \approx \! \xi^{2}_{lf} + 2\xi_{lf}\dxi(t') + \dxi^{2}(t') \,\, , \label{eq:xi_separation}
\eeq
with $\xi_{lf}$ being the quasi-static shift changing between measurements (\textit{i.e.}~coming from the noise spectrum for $\omega_{0} \! < \! \omega \! < \! 1/\te$), and with $\dxi(t')$ being the high-frequency component. The low-frequency cutoff is $\omega_{0} \! \approx \! \text{max}(1/t_{c},1/T_{M})$, with $T_{M}$ being the total data acquisition time, or, in the case of using the gathered data to estimate the slow drift of $\xi^{2}$, by the timescale on which this information is fed back into the averaging procedure (note that such a procedure of correcting for slowly evolving energy offset and suppressing decoherence was simulated in \cite{Falci_PRL05} for the free evolution case). 
Typically $T_{M}$ is orders of magnitude larger than $t$, and in the case of noise with finite $t_{c}$ determining the low-frequency cutoff, we assume here $t_{c} \! \gg \! \te$. In both cases  we have  $\mean{\xi^{2}_{lf}} \! \gg \! \mean{\dxi^{2}}$ when $\beta \! > \! 1$, and the dominant noisy term is $2\xi_{lf}\dxi(t')$ (note that the influence of the quasi-static shift $\xi^{2}_{lf}$ is simply removed by the DD sequence when the  Hamiltonian approach from Eq.~(\ref{eq:H}) is used, as is the case in this Section).  
This amounts to an observation that in the presence of $1/f^{\beta}$ noise, for $T_{M}$, $t_{c} \! \gg \te$ we average over evolutions of qubits operated \textit{in the neighborhood} of the OWP: the quasi-static fluctuations $\xi_{lf}$ are shifting the qubit away from the ``true'' OWP, and they renormalize the fast noise $\dxi(t')$ affecting the qubit, so that the dephasing is caused by the $2\xi_{lf}\dxi(t')$ term. This crucial aspect of dephasing due to quadratic coupling to noise with a significant low-frequency component was discussed in Ref.~\cite{Bergli_PRB06}, where $\xi(t')$ being a sum of many random telegraph signals was considered, and the slowest fluctuators were shown to determine the effect that the fast fluctuators have on the qubit.

Let us look at noise with power concentrated at very low frequencies, i.e.~in the range of $\omega \! < \! 1/\te$. Specifically, let us take $S(\omega) \! = \! A_{\beta}/|\omega|^{\beta}$ with $\beta \! > \! 1$ in this frequency range. At higher $\omega$ the spectrum can be of any shape, as long as the total contribution of the high frequencies to the noise power is much smaller than the low-frequency contribution. As mentioned in Sec.~\ref{sec:short}, $S_{2}(\omega)$ is dominated by regions where $S(\omega)$ diverges, and taking only them into account we obtain  for $\omega\! \gg \! \omega_{0}$ the following expression: 
\beq
S_{2}(\omega) \approx \frac{4}{\pi(\beta-1)} \frac{A_{\beta}}{\omega^{\beta-1}_{0}} S(\omega) =  4\sigma^{2}_{0}S(\omega) \,\, , \label{eq:S2beta}
\eeq 
where $\sigma^{2}_{0}$ is the standard deviation of the low-frequency ($\omega\! < \! 1/\te$) noise,
\beq
\sigma^{2}_{0} \! = \! \int_{\omega_{0}}^{1/\te} S(\omega)\text{d}\omega/\pi \,\, . \label{eq:s0}
\eeq 
It is also useful to rederive these results starting from the correlation function of $\xi^{2}(t')$ noise while using the above-discussed separation of $\xi(t')$ into a quasi-static part $\xi_{lf}$ and the ``fast'' noise $\dxi(t')$:
\begin{align}
C_{\xi^2}(t') & \equiv \mean{\xi^{2}(t')\xi^2(0)} - \mean{\xi^2}^2 = 2\mean{\xi(t')\xi(0)}^2 \nonumber \\
& \approx 2\mean{\dxi(t')\dxi(0)}^2 + 4 \sigma^{2}_{0} \mean{\dxi(t')\dxi(0)} + 2\sigma^{4}_{0} \,\, ,
\end{align}
where we identified $\sigma^{2}_{0}$ with $\mean{\xi_{lf}^2}$. After neglecting the first term (because $\mean{\dxi^2}\! \ll \! \sigma^{2}_{0}$) the Fourier transform of the above expression evaluated at finite $\omega$ agrees with Eq.~(\ref{eq:S2beta}).

We can immediately see now how the high- and low-frequency noise components work together by looking at the lowest  order linked term, $R_{2}(\te)$, which is given by Eq.~(\ref{eq:R2}). Through the presence of $S_{2}(\omega)$ in that formula $R_{2}$ is affected both by high-$\omega$ fluctuations, since the filter $|\tilde{f}_{t}(\omega)|^{2}$ picks out fluctuations with $\omega \! \approx \! n\pi/\te$, and by very low frequencies, since $S_{2}(\omega)$ depends on $\omega_{0}$ through $\sigma^{2}_{0}$.

The essence of the calculation below is separate averaging over the slow and fast fluctuations.
Splitting the noise into a quasi-static part and a high-frequency part, $\xi(t') \! = \! \xi_{lf} + \dxi(t')$, we have
\beq
W(\te) = \left \langle   \exp\left( -i\vq \int f_{t}(t') [ \dxi^{2}(t') + 2\xi_{lf}\dxi(t')]\text{d}t' \right )   \right \rangle \,\, .
\eeq
Now we perform the average over the low frequencies. Using the fact that $\xi_{lf}$ is Gaussian distributed with standard deviation $\sigma^{2}_{0}$, we arrive at an expression to be averaged only over the high frequencies (hf) 
\begin{eqnarray}
W(\te) & = & \Big \langle \exp \big [ -i\vq\int f_{t}(t')   \dxi^{2}(t')\text{d}t' +  \nonumber \\
& & \!\!\!\!\!\!\!\!\!\!\!\! -2\sigma^{2}_{0}v^{2}_{2} \int \text{d} t_{1} \int \text{d} t_{2} f_{t}(t_{1}) f_{t}(t_{2}) \dxi(t_{1})\dxi(t_{2})  
\big ]  \Big \rangle_{\text{hf}} \,\, .  \label{eq:Wfull}
\end{eqnarray}
In Eq.~(\ref{eq:Wfull}) the second term is expected to dominate when $\sigma^{2}_{0} \! \gg \! \mean{\dxi^{2}}_{\text{hf}}$, i.e.~when $\text{min}(T_{M},t_{c}) \! \gg \! \te$. 
The calculation of the average involving only this term can be done using a linked cluster expansion similar to the one previously discussed. One only has to be careful with the definition of linked and unlinked clusters when expanding in powers of the second term, $\mathcal{X}(\te)$, in Eq.~(\ref{eq:Wfull}): the correct definition of unlinked cluster of $k$-th order is that it can be written as a product of terms which contain integrals over sets of time variables $\{t_{k}\}$ coming from \emph{disjoint} sets of $\mathcal{X}$ forming $\mathcal{X}^{k}$. 
However, the same result can be obtained in a simpler way by coming back to Eq.~(\ref{eq:Rt}), into which we plug $C(t) \! = \! \mean{\dxi(t)\dxi(0)}_{\text{hf}} + \sigma^{2}_{0}$, and keep only the terms with the maximal power of $\sigma_{0}$, \textit{i.e.}~the ones in which every second $C(t_{kl})$ is replaced by $\sigma^{2}_{0}$. Only terms with even $k$ survive then, and 
\beq
R_{2n}(\te) \approx  [R^{l}_{2}(\te)]^{n}(2\sigma_{0})^{2n}\,\, , \label{eq:R2n} 
\eeq
where $R^{l}_{2}(\te)$ is given in Eq.~(\ref{eq:R2ldef}). Equivalently we have
\beq
\chi_{2n}(\te) \approx -\frac{1}{2}\frac{(-1)^{n}}{n}\left[ 2\chi_{2}(\te) \right ] ^{n} \,\, ,\label{eq:chi2n}
\eeq
which means, for example, that $|\chi_{4}(t)| \! = \! \chi^{2}_{2}(t)$.
The quality of this approximation is illustrated in Fig.~(\ref{fig:R2R4}), where one can see how the exact calculation of $\chi_{4}(\te)$ for OU noise agrees with the above formula when $t \! \ll \! t_{c}$. Using Eq.~(\ref{eq:chi2n}) we can now write out all the terms in the exponent in Eq.~(\ref{eq:WR}). After noticing that the appearing sum is in fact an expansion of the $-\frac{1}{2}\ln(1+x)$ function with $x\! \equiv \! 4\vq^2\sigma^{2}_{0}R^{l}_{2}(\te)$ we arrive at the final expression for the decoherence function: 
\beq
W(\te) \! = \! \frac{1}{\sqrt{1+4\vq^2\sigma^{2}_{0}R^{l}_{2}(\te)}} = \frac{1}{\sqrt{1+2\chi_{2}(t)}} \,\, .  \label{eq:Wlf}
\eeq
This is the main result of this paper for the case of noise with a dominant low-frequency component. 
For large $n$ we can use Eq.~(\ref{eq:Rl}) to relate $R^{l}_{2}(\te)$ to $S(n\pi/\te)$. For $S(\omega  \! \approx \! n\pi/\te) \! \propto \! 1/\omega^{\beta}$ we have
\beq
W(\te) \approx \frac{1}{\sqrt{2}}\left( \frac{T_{2}}{\te} \right )^{\frac{\beta+1}{2}} \,\, \text{for} \,\, \te \! \gg \! T_{2} \,\, ,\label{eq:powerlaw}
\eeq
where the characteristic decay timescale fulfills 
\beq
T_{2} \sim n^\gamma / T^{\eta}_{M} \,\,\,  \text{where} \,\, \gamma\! = \! \frac{\beta}{\beta+1} \,\,  \text{and} \,\,  \eta \! = \! \frac{\beta-1}{\beta+1} \,\, .   \label{eq:powerlaws}
\eeq
Note that if at high frequencies we have $S(\omega \! \approx \! n\pi/\te) \! \propto 1/\omega^{\beta'}$ (while the low-frequency spectrum is still described by exponent $\beta$), in the above formulas $\beta'$ replaces $\beta$ in all the places with the exception of the numerator of the formula for $\eta$. Clearly the key signature of $1/f^{\beta}$ type noise dominating the decoherence at an OWP is the power-law asymptotic decay of $W(\te)$.
Similar tails were obtained for free evolution  \cite{Makhlin_PRL04,Falci_PRL05} and spin echo \cite{Ithier_PRB05} at the OWP for the case of quasi-static noise (i.e.~noise with high-frequency cutoff $\omega_{\text{h}}$ leading to complete loss of phase coherence on timescale $\te\! \ll \! 1/\omega_{\text{h}}$). Equation (\ref{eq:Wlf}) generalizes these results to the case of decoherence under the influence of multiple pulses, with the only assumption being that $\sigma_{0}^{2}$ is closely approximating the total power of the noise. Another signature of dephasing induced by quadratic coupling to noise with strong low-frequency spectral component is the dependence of the measured coherence signal on the total measurement time $T_{M}$. For linear coupling to $1/f^{\beta}$-type noise such a dependence appears only in the case of free evolution (without pulses) \cite{Makhlin_CP04,Ithier_PRB05}. At the OWP this effect is present also under dynamical decoupling.

\begin{figure}[t]
\includegraphics[width=\linewidth]{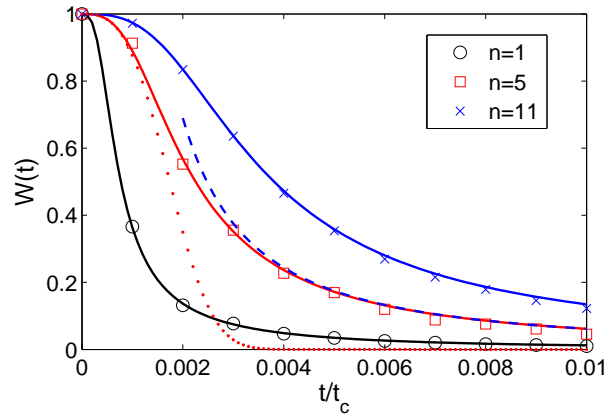}
\caption{(Color online)~Decoherence due to OU noise (with $\sigma\! = \! 1$) at an OWP for CP sequence with $n\! = \! 1$, $5$, and  $11$. Symbols are the results of numerical simulation. For each  $\te$ the averaging time was $T_{M} \! = \! M\te$ with $M\! = \! 10^{6}$, so that the resulting $\sigma^{2}_{0}$ was well approximated by the total power of the OU noise, $\sigma^{2}$. With coupling $v_{2} \! = \! 10^{5}/t_{c}$, the coherence decay in the presented time range is due to the $1/\omega^{2}$ tail of $S(\omega)$. 
The solid lines are obtained using Eq.~(\ref{eq:Wlf}). For $n\!= \! 5$  the dotted line is the Gaussian approximation, and the dashed line is $W(\te)\! \sim \! \te^{-3/2}$ asymptotics from Eq.~(\ref{eq:powerlaw}).
}
 \label{fig:Wlf}
\end{figure}

When $W(\te)$ is measured reliably for large $n$ in a wide range of $\te$, then Eq.~(\ref{eq:Wlf}) together with Eq.~(\ref{eq:Rl}) can be used to reconstruct $\vq^{2}\sigma^{2}_{0}S(\omega)$. Let me remind the reader that Eq.~(\ref{eq:Wlf}) holds for $S(\omega  \! \approx \! n\pi/\te)$ being of any form, we only require the low-frequency part to be diverging as $1/\omega^{\beta}$ and dominating the total noise power.
Alternatively, when the noise is of $1/\omega^{\beta}$ form also at high frequencies, fitting of the $n$ dependence of the power-law tail of $W(\te)$ using Eq.~(\ref{eq:powerlaws}) allows for inferring the value of $\beta$. If the noise is of $1/\omega^{\beta}$ only at low frequencies, then the dependence of $W(\te)$ on $T_{M}$ can be used to fit the value of $\beta$.

In Fig.~\ref{fig:Wlf} Eq.~(\ref{eq:Wlf}) is compared with the results of numerical simulations of dephasing due to noise with $S(\omega) \! \propto \! 1/\omega^{2}$ and a low-frequency cutoff at $\omega_{0} \! \ll \! 1/\te$ (actually an OU noise strongly coupled to the qubit causing dephasing for $\te \! \ll \! t_{c} \!=\! \omega_{0}^{-1}$). One can also see there how the Gaussian approximation fails when $W(\te)$ decays significantly on timescale of $\te \! \ll \! t_{c}$. The asymptotic decay of coherence, according to Eq.~(\ref{eq:powerlaws}), should be given in this case by $1/t^{3/2}$, and the dashed line for $n\! = \! 5$ shows that it is indeed a good approximation to the results of the numerical simulation.

The above derivation is based on the assumption that the contribution of the first term in the exponent in Eq.~(\ref{eq:Wfull}) can be neglected, i.e.~when $1/\omega_{0} \! \gg \! \te$. This in fact is only a necessary condition.
Let us discuss now the approximate sufficient condition. In the second order with respect to this term we obtain $\chi_{2}^{\text{hf}}(\te)$, which for $1/\omega^\beta$ noise is given by a formula analogous to the one for $\chi_{2}(\te)$, only with $\sigma^{2}_{0}$ replaced by
\beq
\sigma_{\te}^{2} \approx \int_{1/\te}^{\infty} S(\omega)\frac{\text{d}\omega}{\pi} \,\, ,
\eeq 
so that we have $\chi^{\text{hf}}_{2}(\te)/ \chi_{2}(\te) \! = \! \sigma^{2}_{\te}/\sigma^{2}_{0}$. For noise with low-frequency cutoff determined by the measurement time $T_{M}$ this means that $\chi^{\text{hf}}_{2}(\te)/ \chi_{2}(\te) \! = \! (\te/T_{M})^{\beta -1}$. Using $\chi_{2}(\te) \! =\! (t/T_{2})^{\beta+1}$ we see that $\chi_{2}^{\text{hf}}(\te) \! \ll \! 1$ when $(t/T_{M}) \! \ll \! (T_{2}/T_{M})^{1/2+1/2\beta}$. 
In the case of OU noise with low-frequency cutoff given by $1/t_{c}$, which is used in simulations shown in the Figures, this means that for the high-frequency correction to be negligible we need $t/t_{c} \! \ll \! (T_{2}/t_{c})^{3/4}$. With $T_{2}$ values in Fig.~\ref{fig:Wlf} this mean that $t/t_{c}$ should not be larger than about 10 times $T_{2}/t_{c}$. In fact, at longer time $\te$, when the previously derived $W(\te) \! \approx \! 1/\sqrt{2\chi_{2}(\te)} \! \ll \! 1$, in order to have $\chi^{\text{hf}}_{2}(\te) \! \ll \! 1$ we need
\beq
\frac{1}{\pi} \frac{\te}{t_{c}} \ll W^{2}(\te) \,\, .
\eeq 
This means that in order for the formulas from this section to work down to $W(\te)\! \approx \! 0.1$ in the case of noise used in the simulations shown in Fig.~\ref{fig:Wlf}, the condition $\te/t_{c} \! \ll \! \pi\cdot 10^{-2}$ has to be fulfilled. The results of numerical simulations in Fig.~\ref{fig:Wlf} are shown for $\te/t_{c} \! \leq \! 10^{-2}$, and in fact at longer $\te$ the simulated $W(\te)$ falls below the results of Eq.~(\ref{eq:Wlf}) due to the influence of high-frequency noise components. The fact that simulation datapoints are slightly below the theoretical solid line in Fig.~\ref{fig:Wlf} at $\te/t_{c} \! \approx \! 10^{-2}$ is the precursor of this.

\section{Simulations for coupling to transverse noise} \label{sec:transverse}
Let us come back now to the issue of the relation between the calculations using an effective Hamiltonian from Eq.~(\ref{eq:H}) (i.e.~all the results presented before) and the behavior of coherence at the extrinsic OWP, appearing at large qubit splitting $\Omega$ in the presence of transverse noise. We thus take the Hamiltonian from Eq.~(\ref{eq:Ht}), describing an often-encountered case of uniaxial transverse noise, and we set out to re-analyze the dynamics of coherence decay under dynamical decoupling. 

\begin{figure}[t]
\includegraphics[width=\linewidth]{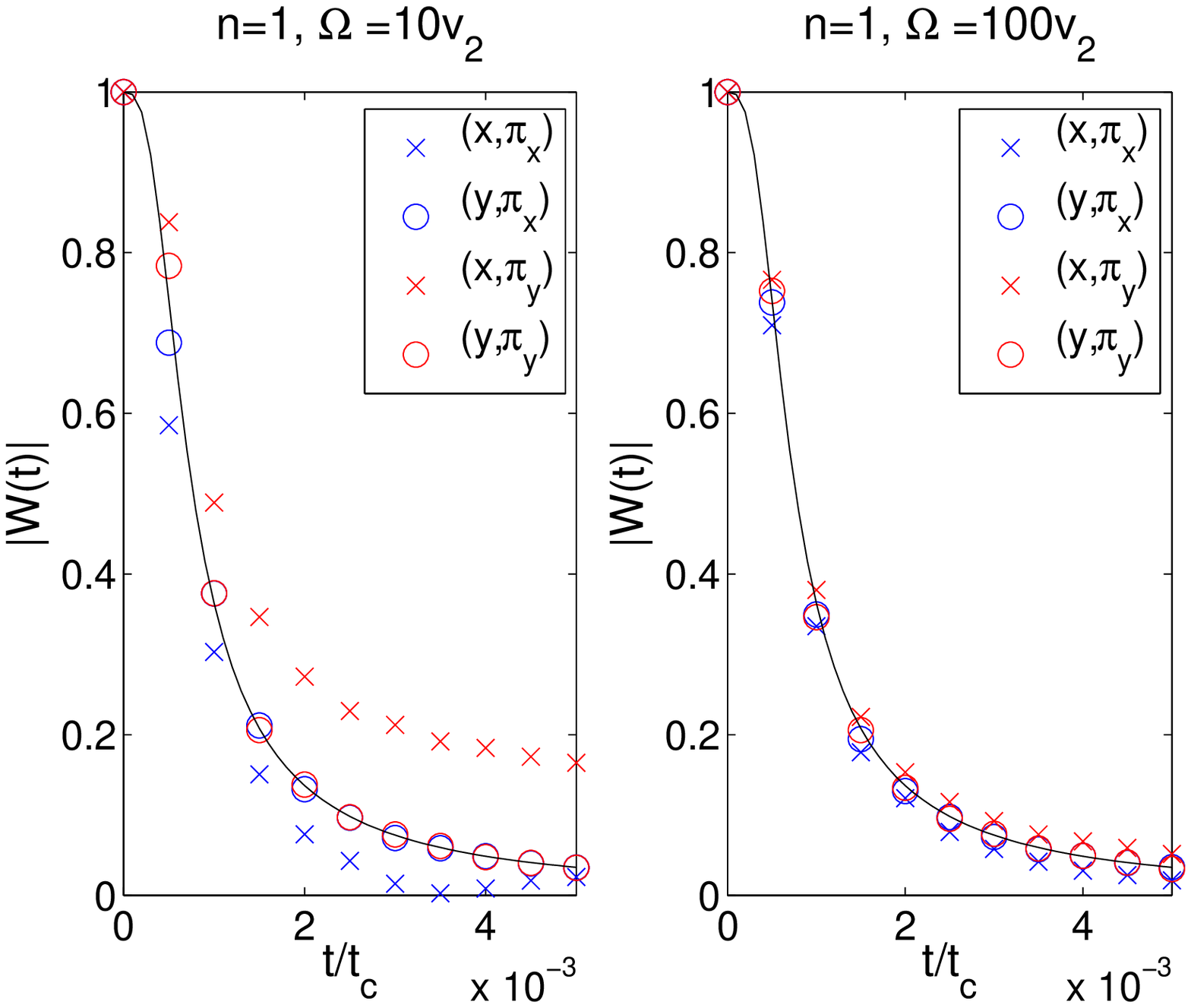}
\caption{(Color online)~Decoherence due to transverse OU noise with $\sigma \! = \! 1$ for CP sequence with $n\! = \! 1$ (spin echo). The transverse coupling constant is $v_{t}\! = \! \sqrt{2\Omega v_{2}}$ and $v_{2}\!= \! 10^{5}/t_{c}$. The notation $(i,\pi_{j})$ specifies the initial direction of the qubit ($i\! =\! x$, $y$) and the axis about which the $\pi$ pulse is applied ($j\!= \! x$, $y$).
The solid lines are obtained using Eq.~(\ref{eq:Wlf}). 
}
 \label{fig:SE_pulses}

\includegraphics[width=\linewidth]{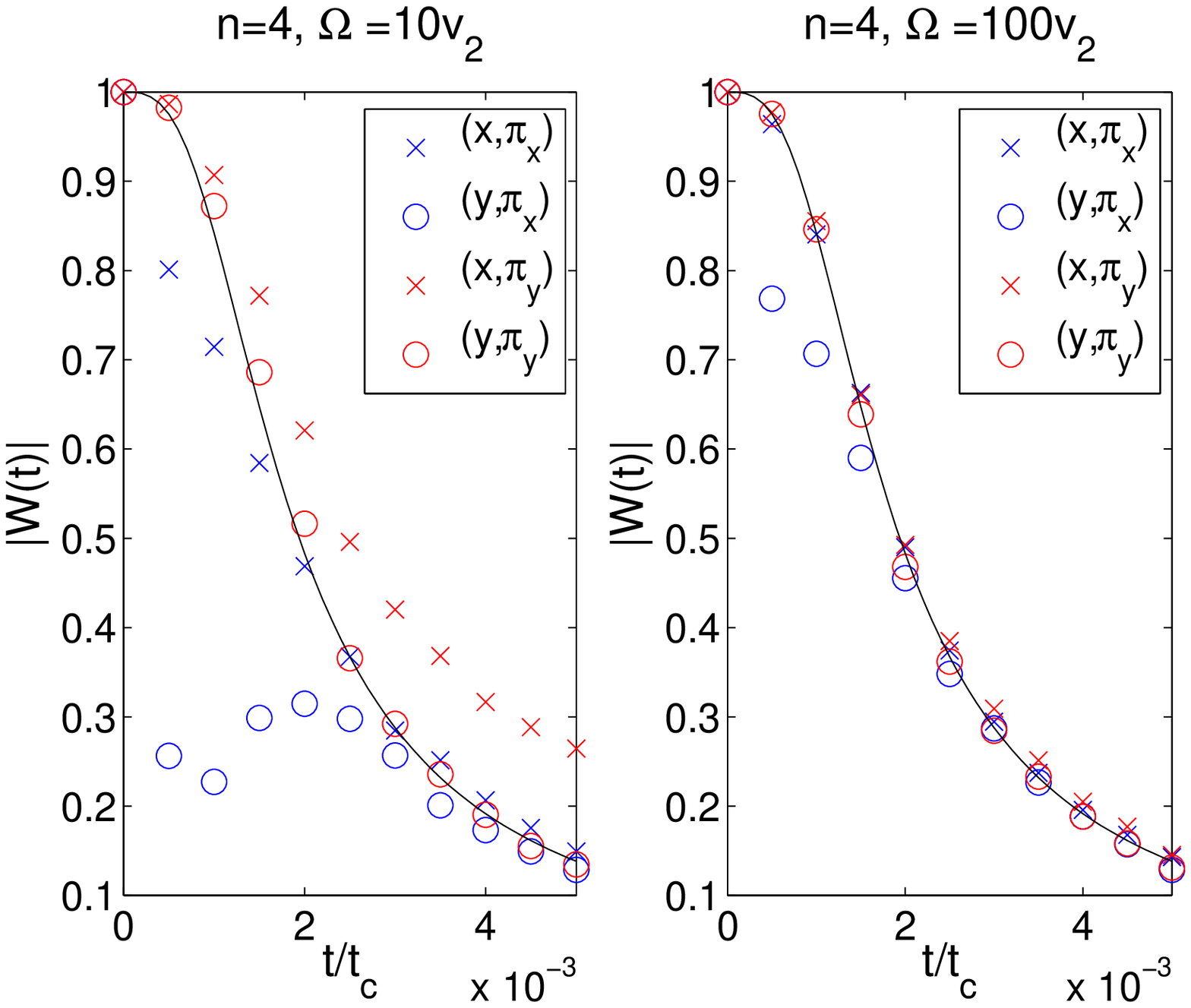}
\caption{(Color online)~Decoherence due to transverse OU noise with $\sigma \! = \! 1$ for CP sequence with $n\! = \! 4$. Parameters and symbols are the same as in Fig.~\ref{fig:SE_pulses}.
}
 \label{fig:CP4_pulses}
\end{figure}

We focus on the case in which the qubit is initialized along the $x$ or $y$ axis (i.e.~in a superposition of eigenstates of $\Omega \hat{\sigma}_{z}/2$), and the coherence of the qubit (the off-diagonal element of its density matrix in the same basis) is monitored as a function of time.
The dephasing due to transverse noise can be approximately mapped on pure dephasing described by Eq.~(\ref{eq:H}) when $\Omega$ is much larger than the rms amplitude of the noise felt by the qubit, $v_{t}\sigma$. In this regime, as discussed in Section \ref{sec:OWP}, in the lowest order in transverse perturbation we can derive Eq.~(\ref{eq:H}) from Eq.~(\ref{eq:Ht}). For any $\Omega$ present in the transverse coupling Hamiltonian, the effective pure-dephasing Hamiltonian has $v_{2} = v^{2}_{t}/2\Omega$. Now we want to give at least a qualitative answer to the question: for what values of $\Omega$, $n$, and $t$, the previously obtained results correspond also to dephasing caused by transverse noise.

The first thing that should be noted is that now the influence of ideal (error-free) $\pi$ pulses does depend on the axis with respect to which these rotations are performed. As noted in \cite{Bergli_PRB07}, the $\pi$ pulse about the axis along which the noise couples to the qubit (the $x$ axis in Hamiltonian (\ref{eq:Ht})) should be less efficient in mitigation of noise-induced dephasing compared to the pulse about a perpendicular $y$ axis. This is easiest to see in the case of quasi-static noise, for which a single $\pi_{y}$ pulse leads to a perfect echo of the coherence (more precisely, an echo sequence with such a pulse gives $\mean{\hat{\sigma}_{x}(t)} \! = \! -\mean{\hat{\sigma}_{x}(0)}$ and $\mean{\hat{\sigma}_{y}(t)} \! = \! \mean{\hat{\sigma}_{y}(0)}$ at the echo time $t$), while the $\pi_{x}$ pulse gives only a partial recovery of coherence. A strong asymmetry between performance of $\pi_{x}$ and $\pi_{y}$ pulses was also seen in the case of coupling to non-Gaussian random telegraph noise \cite{Bergli_PRB07,Ramon_PRB12}. The application of $\pi_{z}$ pulses (which in the currently considered case are expected to have some influence, since the noise couples to the qubit through $\hat{\sigma}_{x}$) requires the interpulse time to be $\ll 1/\Omega$, otherwise the dephasing is actually enhanced \cite{Faoro_PRL04}. For this reason only $\pi_{x}$ and $\pi_{y}$ pulses are considered below.

\begin{figure}[t]
\includegraphics[width=\linewidth]{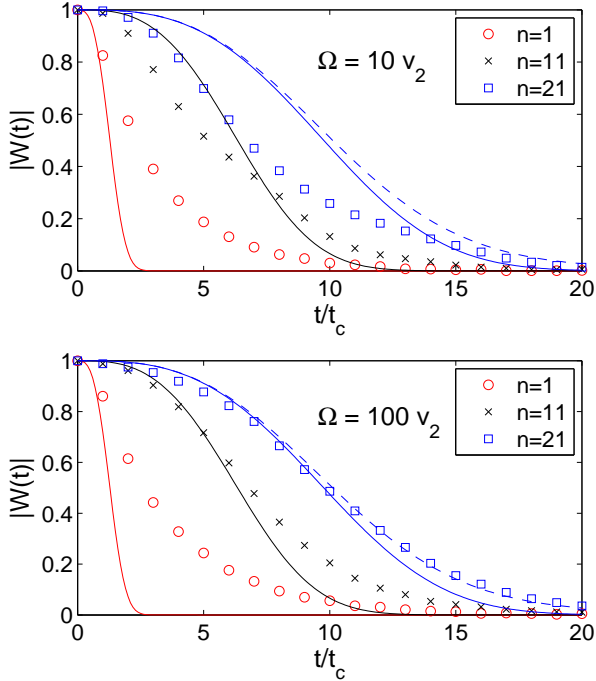}
\caption{(Color online)~Decoherence due to transverse OU noise (with $\sigma\! = \! 1$) at an OWP for CPMG $(y,\pi_{y})$ sequence with $n\! = \! 1$, $11$, and  $21$. Symbols are the results of numerical simulation for transverse noise with $v_t$ couplings chosen such that for qubit splitting $\Omega\!=\! 10 v_{2}\sigma^{2}$ (upper panel) and $\Omega\!=\! 100 v_{2}\sigma^{2}$ (lower panel) the results are expected to match the effective Hamiltonian results from Fig.~(\ref{fig:gaussian}), where $v_{2}\! = \! 1/t_{c}$ was used.  Solid lines are the Gaussian approximation $W(\te)\! =\! \exp[-\chi_{2}(t)]$, while the dashed line for $n\!= \! 21$ is $W(\te)\! =\! \exp[-\chi_{2}(t) - \chi_{4}(t)]$.      
}
 \label{fig:Wg_transverse}
\end{figure}

\begin{figure}[t]
\includegraphics[width=\linewidth]{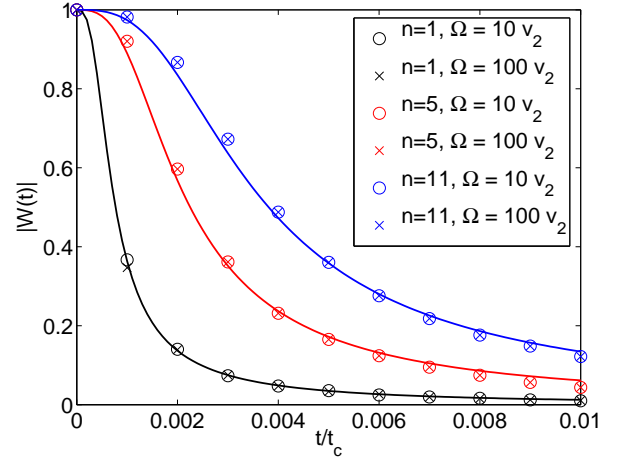}
\caption{(Color online)~Decoherence due to transverse OU noise for CPMG  $(y,\pi_{y})$ sequence with $n\! = \! 1$, $5$, and  $11$. Symbols are the results of numerical simulation for transverse noise with $v_t$ coupling chosen such that the results are expected to match the effective Hamiltonian calculations from Fig.~\ref{fig:Wlf} (where $\vq \! = \! 10^{5}/t_{c}$ has been used). The circle and cross symbols correspond to two values of $\Omega$.
The solid lines are obtained using Eq.~(\ref{eq:Wlf}). 
}
 \label{fig:Wlf_transverse}
\end{figure}

However, in the case of coupling to transverse Gaussian noise which is not exactly quasi-static, these differences are less pronounced. In Fig.~\ref{fig:SE_pulses} the results of simulations of the echo sequence are shown for the case of Ornstein-Uhlenbeck noise, in the coupling regime in which the coherence signal decays for $t \! \ll \! t_{c}$. Two possible initial states $i\! = \! x$, $y$ of the qubit (along the $x$ and $y$ axes) and two above-discussed $\pi_{j}$ pulses are considered -  the notation $(i,\pi_{j})$ is used to label the results.
For $\Omega \! = \! 10\vq$ (and assuming $\sigma\! = \! 1$ for the noise) we see that the coherence is best protected in the $(x,\pi_{y})$ case, while $(x,\pi_{x})$ fares the worst. The differences between various cases almost disappear for $\Omega \! = \! 100\vq$ (this corresponds to $v_{t} \! = \! \sqrt{200}\vq \! \ll \! \Omega$). It is worth noting that the $(y,\pi_{y})$ results are matching the results of the effective Hamiltonian calculation (the solid line) very well even for the lower value of $\Omega$.

In Fig.~\ref{fig:CP4_pulses} analogous results are shown for the CP sequence with $n\! = \! 4$ pulses. The accumulation of errors (leading to apparent coherence loss) is evident, especially in the $(y,\pi_{x})$ case (which was somewhat surprisingly agreeing very well with the effective Hamiltonian results for $n\! = \! 1$). However, $(y,\pi_{y})$ results are still matching the previously described theory, even at quite low values of $\Omega$. In the following we will then use the transverse noise simulations for this case to check whether the previously presented results correspond also to the case of an extrinsic OWP.

It is interesting to note that the $(y,\pi_{y})$ case corresponds precisely to the Car-Purcell-Meiboom-Gill sequence \cite{Meiboom_Gill}: the qubit initialized along the $z$ axis is rotated by $\pi/2$ with a pulse about the $x$ axis, and subsequent $\pi$ pulses are about the $y$ axis. This sequence is known to be robust against systematic pulse errors \cite{Borneman_JMR10}, which is probably related to the above-discussed behavior - slow components of  the noise cause quasi-static perturbations of qubit's quantization axis, which translate into quasi-systematic pulse errors. 

In Figures \ref{fig:Wg_transverse} and \ref{fig:Wlf_transverse} I show the results of simulations of decoherence due to transverse noise, with parameters chosen to match the parameters of previously considered effective Hamiltonians. These Figures should be compared with previously presented Figs.~\ref{fig:gaussian} and \ref{fig:Wlf}, respectively. 

Comparison of Fig.~\ref{fig:Wg_transverse} with Fig.~\ref{fig:gaussian} shows that the effective ``Gaussianization'' of noise discussed in Section \ref{sec:Gaussian} holds quite well in the case of an extrinsic OWP, provided that $\Omega$ is large enough - the results for $\Omega \! = \! 10\vq$ visibly differ from those from Fig.~(\ref{fig:gaussian}), but when using $\Omega \! = \! 100 \vq$ the differences are almost invisible. Actually they are most pronounced at short times, at which the apparent decoherence due to the influence of pulses pushes the simulation results below the theoretical result - note that the points for $n\! =\! 21$ and $\Omega\! = \! 100 \vq$ are slightly below the theoretical solid line at short times. 

On the other hand, when decoherence occurs at times $t \! \ll \! t_c$, the simulations of the influence of transverse noise agree very well with the analytical results from Section \ref{sec:1f}, even at smaller values of $\Omega$, see Fig.~\ref{fig:Wlf_transverse}. The results shown in this Figure for $\Omega\! = \! 10 \vq$ and $100 \vq$ are indistinguishable from those presented earlier in Fig.~\ref{fig:Wlf}.

\section{Conclusion} \label{sec:discussion}
In this paper I have discussed how one can perform the spectroscopy of the environmental (classical and Gaussian) noise $\xi(t')$, when the stochastic contribution to the qubit's energy splitting is $\propto \xi^{2}(t')$. Furthermore, the numerical simulations have shown that coherence dynamics under dynamical decoupling for a qubit affected by purely transverse noise, is described by the same analytical theory quite well (at least for moderate numbers of pulses used in this paper) when the CPMG pulse sequence is used. 
For noise with no low-frequency divergence of its spectral density $S(\omega)$ (i.e.~for noise with well-defined correlation time comparable or shorter than the qubit's evolution timescale), using a large number of pulses $n$ one can use established methods of dynamical decoupling noise spectroscopy to reconstruct the spectral density of $\xi^{2}(t')$ process. For noise with $S(\omega) \! \sim \! 1/\omega^{\beta}$ at low frequencies (and with $\beta \! > \! 1$ guaranteeing that most of the noise power is concentrated at low frequencies), a closed formula for the qubit's dephasing under dynamical decoupling has been derived. At large $n$, the coherence at time $\te$ is determined by $S(n\pi/\te)$, as in the case of the linear coupling to noise, but the functional form of the coherence decay is distinct from that case: the decay at long times is not superexponential, but of power-law character. Furthermore, the coherence signal depends on the low-frequency cutoff of the noise spectrum, which is typically given by an inverse of the total data acquisition time  $T_{M}$. Because of this, the information about both the low- and high-frequency part of $S(\omega)$ can be obtained from measurements of the characteristic decay timescale and of the asymptotic behavior of coherence.
The presented results therefore show how one can obtain quantitative information on the noise spectrum by measuring decoherence of the qubit operated at its optimal point.

\section*{Acknowledgements}
I would like to thank C.~M.~Marcus for inspiring discussions. I also thank V.~Mkhitaryan, V.~V.~Dobrovitski, and G.~Ramon for discussions and comments. 
This work is supported by funds of Polish National Science Center (NCN), grant no.~DEC-2012/07/B/ST3/03616.

\end{document}